\documentstyle[multicol,prl,aps,epsfig]{revtex}
\def\Journal#1#2#3#4{{#1} {\bf #2}, #3 (#4)}

\def\NIMA{{ Nucl. Instrum. Methods} A}
\def\NPA{{ Nucl. Phys.} A}
\def\NPB{{ Nucl. Phys.} B}
\def\NPBproc{{ Nucl. Phys.} B(Proc. Suppl.)}

\def\PLB{{ Phys. Lett.}  B}
\def\PRL{ Phys. Rev. Lett.}
\def\PRD{{ Phys. Rev.} D}

\def\EUPJ{{ Eur. Phys. J.} C} 
\def\JPG{{ J. Phys. G:} Nucl. Part. Phys.}

\newcommand{\etal}{{\it et al.}}
\begin{document}

\title{Single-spin Azimuthal Asymmetries in 
Electroproduction of Neutral Pions in Semi-inclusive Deep-inelastic 
Scattering}
%
\author{ 
A.~Airapetian,$^{31}$
N.~Akopov,$^{31}$
Z.~Akopov,$^{31}$
M.~Amarian,$^{26,31}$
E.C.~Aschenauer,$^{7}$
H.~Avakian,$^{11}$
R.~Avakian,$^{31}$
A.~Avetissian,$^{31}$
E.~Avetissian,$^{31}$
P.~Bailey,$^{15}$
% ...
\medskip\\ \centerline{(The HERMES Collaboration)}\medskip
}

\address{ 
% ...
$^7$DESY Zeuthen, 15738 Zeuthen, Germany\\
% ...
$^{11}$Istituto Nazionale di Fisica Nucleare, Laboratori Nazionali di Frascati, 00044 Frascati, Italy\\
% ...
$^{15}$Department of Physics, University of Illinois, Urbana, Illinois 61801, USA\\
% ...
$^{26}$Istituto Nazionale di Fisica Nucleare, Sezione Sanit\`a and Physics Laboratory, Istituto Superiore di Sanit\`a, 00161 Roma, Italy\\
% ...
$^{31}$Yerevan Physics Institute, 375036, Yerevan, Armenia\\
}

%
 \maketitle
%
\begin{abstract}
A single-spin asymmetry in the azimuthal distribution of neutral pions
 relative to the lepton scattering plane
has been measured for the first time
in deep-inelastic scattering of positrons off longitudinally polarized protons.
The analysing power in the $\sin \phi$ moment of the cross section is 
$0.019 \pm 0.007(\mbox{stat.}) \pm 0.003(\mbox{syst.})$. 
This result is compared to single-spin asymmetries for charged pion production
measured in the same kinematic range. 
The $\pi ^0$ asymmetry is of the same size as the
$\pi ^+$ asymmetry and shows a similar dependence on the relevant kinematic 
variables.
The asymmetry is described by a phenomenological calculation based on a  
fragmentation function that represents sensitivity to the transverse 
polarization of the struck quark.
\end{abstract}
\begin{figure}[htb]
\hspace*{-0.2cm}\epsfxsize 17.0 cm {\epsfbox{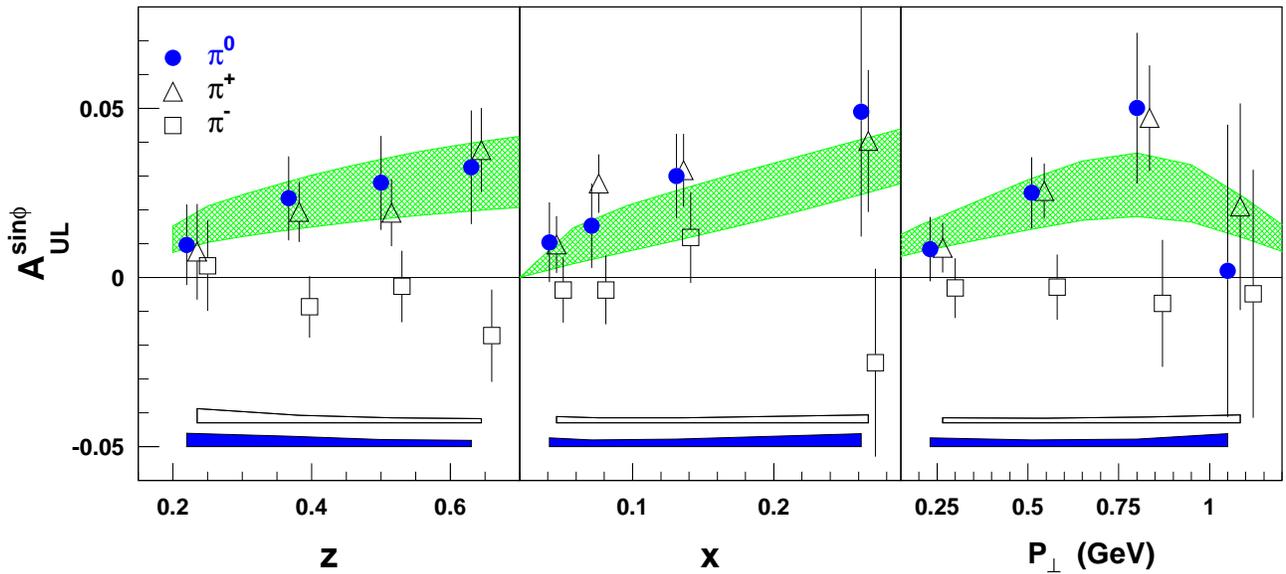}}
\caption{
Analysing power in the $\sin \phi$ moment  for $\pi ^0$ (circles) 
compared to previous results~\protect\cite{hermes_charged_pion_ssa:1999} 
for $\pi ^+$ (triangles) and $\pi ^-$ (squares) 
as a function of the pion fractional energy $z$, the Bjorken variable $x$, and of 
the pion transverse momentum $P_\perp$. 
Error bars include the statistical uncertainties only. The filled and open bands at 
the bottom of  the panels represent the  systematic uncertainties for neutral and 
charged pions, respectively.
The data for charged pion production are slightly shifted in $z$, $x$, 
and $P_{\perp}$ for clarity.
The shaded areas show a range of predictions of a model calculation
~\protect\cite{Karo+:1998,deSanctis:2000} applied to the case of $\pi ^0$ 
electroproduction (see text).
}
\label{fig_ssa_all}
\end{figure}

%
%
%
%
\begin{multicols}{2}[]
Semi-inclusive pion production in polarized deep-inelastic scattering is a powerful
tool for investigating the spin structure of the nucleon and
providing information about the spin-dependent parton distribution and
fragmentation functions.      
The recent observation of a significant azimuthal asymmetry for semi-inclusive 
$\pi^+$ production in deep-inelastic scattering (DIS) 
of unpolarized positrons off longitudinally polarized protons 
\cite{hermes_charged_pion_ssa:1999} has revealed effects of quark
distribution and fragmentation functions that describe the transverse polarization
of quarks.
In particular, the occurrence of single-spin asymmetries, where only the 
target is polarized,  offers access to the fundamental but still unmeasured
chiral-odd transversity distribution functions 
\cite{Ralston-Soper:1979,Jaffe-Ji:1992} through the so-called Collins effect 
\cite{Collins:1993,Kotzinian:1995,Mulders+:1996,Jaffe+:1998}. 
This effect involves a chiral-odd fragmentation function, the Collins function, 
that describes the fragmentation of a transversely polarized quark into an
unpolarized hadron
and is discussed in recent experimental \cite{hermes_charged_pion_ssa:1999,smc_ssa:1999}
and theoretical publications
\cite{Mulders:1997,Karo+:1998,Karo+:1999,Boer:1999,Efremov+:2000,Boglione+:2000,Anselmino+:2000,Jaffe:2000,deSanctis:2000,Schafer-Teryaev:2000,Ma+:2000}
on this subject.
\\ \indent
Single-spin asymmetries in pion production have already been measured 
in proton-proton scattering experiments \cite{pp_experiments}, 
where the $\pi ^0$ asymmetry is found to be similar to the $\pi ^+$ one, 
but with opposite sign as compared with the $\pi ^-$ one.
This isospin dependence could be understood \cite{Anselmino:1995} in terms of 
the difference between the unpolarized cross sections for $\pi ^+$ and $\pi ^-$ 
production.
However, these asymmetries from proton-proton scattering
 may also arise from initial-state interactions, 
which are negligible for semi-inclusive lepton-nucleon scattering processes.
In the latter case any single-spin asymmetry can originate only from a spin 
dependence in the fragmentation of a polarized quark.
In fact such a dependence is assumed in \cite{Anselmino+:2000,Jaffe:2000,Boros+:1998}
where a sizeable $\pi ^0$ single-spin asymmetry is predicted for 
lepton-nucleon scattering.
\\ \indent
This paper reports the first observation of a single-spin azimuthal asymmetry 
in semi-inclusive neutral pion electroproduction.
The relevant kinematic variables of this process in the target rest frame 
are the spacelike squared four-momentum $Q^2$ of the exchanged virtual photon
 with energy $\nu$,  the pion fractional energy $z = E_{\pi} / \nu$, 
the pion transverse momentum $P_{\perp}$, and
the azimuthal angle $\phi$ of the pion around the virtual-photon axis. 
Here $ E_{\pi}$ is the pion energy, $P_{\perp}$ is defined with
respect to the virtual-photon direction and $\phi$ is defined relative to the
lepton scattering plane.
The fractional energy transferred to the proton is given by $y=\nu / E$ and 
the Bjorken scaling variable is defined as $x = Q^2 / 2M\nu$, where
 $E$ is the lepton beam energy and  $M$ is the proton mass.
\\ \indent
The data were collected in 1996 and 1997
using  a longitudinally polarized hydrogen gas target in the 27.57~GeV HERA 
positron storage ring at DESY.
The average target polarization was 0.86 with a fractional 
uncertainty of 5$\%$.
The scattered positron and  the decay photons from the $\pi^0$  
were detected by the HERMES spectrometer {\cite{hermes:spectrometer}}.
Positrons were distinguished from hadrons with an average efficiency of 
99$\%$ at a hadron contamination of less than 1$\%$ using the information 
from an electromagnetic calorimeter, a transition radiation detector, 
a preshower scintillator detector, and a threshold Cerenkov detector.
The kinematic requirements imposed on the scattered positrons were 
$Q^2 > 1$~GeV$^2$, $0.023 < x < 0.4$, $0.2 < y < 0.85$, and
an invariant mass squared of the initial photon-nucleon system $W^2 > 4$~GeV$^2$.  
\\ \indent
Neutral pion identification was provided by the detection of the two photon 
clusters originating from the $\pi ^0$ decay  in the electromagnetic calorimeter 
\cite{hermes-calo:1996}, each with a minimum energy deposition of 1.0~GeV
and without a corresponding charged track.
The reconstructed photon-pair  invariant mass $m_{\gamma \gamma}$ shows a
clear $\pi ^0$ mass peak with a mass  resolution of about 0.012~GeV.
Neutral pions were selected within the invariant mass range 
of $0.10 < m_{\gamma \gamma} < 0.17$~GeV where background contributions from
uncorrelated photons typically amount to 20$\%$.
An upper limit  of $z<0.7$ was used in order to minimise acceptance effects 
and to suppress possible contributions of exclusive processes.
The requirement $P_{\perp} > 0.05$~GeV ensures an accurate measurement of the 
azimuthal \mbox{angle $\phi$}.
\\ \indent
The analysing powers for unpolarized (U) beam and longitudinally (L) polarized 
target are evaluated as 
\begin{equation}
\label{equ-asymmetry}
 A_{\mathrm{UL}}^{W} =
\frac{\frac{L^{\rightarrow}}{L^{\rightarrow}_{\mathrm{P}}}
\sum_{i=1}^{N^{\,\rightarrow}} W (\phi ^{\rightarrow}_i)\, -\,
      \frac{L^{\leftarrow}}{L^{\leftarrow}_{\mathrm{P}}} 
\sum_{i=1}^{N^{\,\leftarrow}} W (\phi ^{\leftarrow}_i)}
     {\frac{1}{2} [N^{\,\rightarrow}\, +\, N^{\,\leftarrow}]}  \; ,
\end{equation}   
using the weighting functions $W(\phi) = \sin \phi$ and $W(\phi) = \sin 2\phi$. 
Here, the superscripts $\rightarrow$ and $\leftarrow$ denote opposite 
target helicity states.
Each summation runs over the number 
$N^{\rightarrow(\leftarrow)}$ of pions selected for each target 
helicity state and is multiplied with the dead-time corrected luminosities
$L^{\rightarrow(\leftarrow)}$ and $L^{\rightarrow(\leftarrow)}_{\mathrm{P}}$, 
the latter being weighted by the target polarization magnitude.
The analysing powers were determined by integrating over the spectrometer 
acceptance in the kinematic variable $y$ with a mean value of 0.57.
Small corrections were applied for cross-contamination between the $\sin \phi$ and 
$\sin 2\phi$ moments due to the spectrometer acceptance, based on a Monte Carlo 
simulation.
Aside from this, the results have been shown to be little affected by the 
limited acceptance of the HERMES spectrometer.
A background contribution from uncorrelated photons was estimated by varying 
the window in the reconstructed invariant mass of the photon pair. 
It was found to be negligible within the statistical accuracy of the data
and was taken into account in the systematic error. 
The primary  contributions to the systematic uncertainty arise from the
target polarization and from the acceptance corrections.
Radiative effects on the unpola\-rized cross section were evaluated and 
a contribution to the asymmetry of less then 0.1$\%$ averaged over the 
full acceptance was found \cite{Akushevich:1999}.
\\ \indent
The analysing power in the azimuthal $\sin \phi$ moment of the $\pi ^0$ 
production cross section, averaged over $z$, $x$, and $P_{\perp}$ with mean 
values of 0.48, 0.09, and 0.44 GeV, respectively, is 
$0.019 \pm 0.007 \mbox{(stat.)} \pm 0.003 \mbox{(syst.)}$. 
This result is consistent with the $\pi^+$ measurements
for which an analysing power of 
$0.022 \pm 0.005 \mbox{(stat.)} \pm 0.003 \mbox{(syst.)}$
was reported \cite{hermes_charged_pion_ssa:1999}.
The observed azimuthal asymmetry implies
a substantial magnitude for the Collins fragmentation function $H_1^\perp$.
The analysing power in the $\sin 2\phi$ moment for the same process, 
calculated using Eq.~(\ref{equ-asymmetry}), is consistent with zero within 
the statistical uncertainty: 
$0.006 \pm 0.007 \mbox{(stat.)} \pm 0.003 \mbox{(syst.)}$.
This is expected from predictions  for the ratio of 
$A_{\mathrm{UL}}^{\sin 2\phi}$ to $A_{\mathrm{UL}}^{\sin \phi}$
\cite{Karo+:1999}, which are small in the valence region for  
the specific kinematic range of relatively low $Q^2$ and moderate $P_{\perp}$
accessible at HERMES. 
\\ \indent
In Fig.~\ref{fig_ssa_all} the analysing power $A_{\mathrm{UL}}^{\sin \phi}$ 
is shown as a function of the pion fractional energy $z$, the Bjorken scaling 
variable $x$, and the pion transverse momentum $P_{\perp}$, 
after averaging over the other two kinematic variables. 
Also shown are results for charged pions \cite{hermes_charged_pion_ssa:1999}
 obtained in the same kinematic range. 
The $\pi ^0$ and $\pi ^+$ asymmetries exhibit a similar behaviour 
in all kinematic variables.
The  $z$ dependence of the $\pi ^0$ asymmetry is consistent with the 
monotonic increase of the $\pi ^+$ asymmetry.
The increase of $A_{\mathrm{UL}}^{\sin \phi}$ with increasing $x$ suggests
that single-spin asymmetries are associated with valence quark contributions.
%
\end{multicols}
\begin{multicols}{2}[]
\noindent
The dependence on $P_{\perp}$ can be related to the dominant kinematic role of 
the intrinsic transverse  momentum of the quark, if $P_{\perp}$ 
remains below the typical hadronic mass of about 1~GeV
\cite{hermes_charged_pion_ssa:1999}.
Also shown in Fig.~\ref{fig_ssa_all} are the predictions of a model 
calculation \cite{Karo+:1998,deSanctis:2000} for the $\pi ^0$ single-spin
azimuthal asymmetry using isospin and charge conjugation invariance
\cite{Anselmino+:2000}.
In the case of longitudinally pola\-rized nucleons, two distribution functions 
enter the calculation. 
One, $h_{1\mathrm{L}}^{\perp}$, is twist-2 and describes the quark 
transverse spin distribution in a longitudinally pola\-rized nucleon,
while $h_{\mathrm{L}}$ includes an interaction dependent twist-3 part.
Both are related to the twist-2 distribution function $h_1$, called transversity,
that describes the quark transverse spin distribution in a transversely 
polarized nucleon, by
$h_{\mathrm{L}} (x) = h_1(x) - \frac{\mathrm{d}}{\mathrm{d} x}  
h_{1\mathrm{L}}^{\perp (1)} (x)$ \cite{Mulders+:1996},
where $h_{1\mathrm{L}}^{\perp (1)}$ 
is the $k_{\mathrm{T}}^2$-moment of $h_{1\mathrm{L}}^{\perp}$
over the intrinsic quark transverse momentum $k_{\mathrm{T}}$.
Assuming a vani\-shing $h_{1\mathrm{L}}^{\perp (1)}$
\cite{deSanctis:2000}, the number of 
unknown distribution functions can be reduced to one: $h_{\mathrm{L}}\simeq h_1$.
The range of predictions, shown in Fig.~\ref{fig_ssa_all}, is obtained 
by varying $h_1$ between the two assumptions $h_1=g_1$ (non-relativistic limit)
 and $h_1=(f_1 + g_1)/2$ (Soffer inequality), 
with the usual polarized and unpolarized distribution functions
$g_1$ and $f_1$, respectively.
In both cases a simple parameterisation for the spin-dependent time-reversal-odd
fragmentation function $H_1^\perp$ was adopted.
The predictions are consistent with the measured $\pi ^0$ azimuthal asymmetries 
and describe the dependences on the kinematic variables. 
These new $\pi ^0$ data provide additional  information also for other  
phenomenological approaches \cite{Boer:1999,Efremov+:2000,Boglione+:2000,Ma+:2000}.
\\ \indent
In summary, a single-spin azimuthal asymmetry for $\pi ^0$ production
has been measured in semi-inclusive deep-inelastic lepton scattering off a 
longitudinally polarized proton target.
The dependence of this asymmetry on the kinematic variables $x$, $z$, and
$P_\perp$ has been investigated.
The results are similar to the previously measured azi\-muthal asymmetry
for $\pi ^+$ electroproduction, while the $\pi ^-$ asymmetry was consistent
with zero.
This finding can be well described by a model calculation where the asymmetry is 
interpreted as the effect of the convolution of a chiral-odd distribution function 
and a time-odd fragmentation function.
The observed single-spin azimuthal asymmetries for neutral and charged pion 
electroproduction are consistent with the assumed $u$-quark dominance in both
distribution and fragmentation functions.
These results provide evidence in  support of the existence of non-zero 
chiral-odd structures that describe the transverse polarization of quarks. 
New data are expected from future HERMES measurements on a
transversely pola\-rized target, which will give direct access to the transversity
\cite{Nowak+:2000}.  
\\ \indent
We thank M. Anselmino, R.L. Jaffe, A.M. Kotzinian, and P.J. Mulders 
for many interesting discussions on this subject. 
We gratefully acknowledge the DESY management for its support, the staffs at DESY, 
and the collabo\-rating institutions for their significant effort, and our funding
agencies for financial support.  
Additional support for this work was provided by the DAAD and INTAS, HCM, TMR 
network contributions from the
European Community, and the U.S Department of Energy.
%

\end{multicols}

\end{document}